# A Scaling Relation of Anomalous Hall Effect in Ferromagnetic Semiconductors and Metals

Tomoteru Fukumura[1 *], Hidemi Toyosaki[1], Kazunori Ueno[1], Masaki Nakano[1], Takashi Yamasaki[1], and Masashi Kawasaki[1,2]

[1]*Institute for Materials Research, Tohoku University, Sendai 980-8577, Japan*

[2]*CREST, Japan Science and Technology Agency (JST), Kawaguchi, Saitama 332-0012, Japan*

A scaling relation of the anomalous Hall effect recently found in a ferromagnetic semiconductor (Ti,Co)$O_2$ is compared with those of various ferromagnetic semiconductors and metals. Many of these compounds with relatively low conductivity $\sigma_{xx} \leq 10^4\ \Omega^{-1}\cdot\mathrm{cm}^{-1}$ are also found to exhibit similar relation: anomalous Hall conductivity $\sigma_{\mathrm{AH}}$ approximately scales as $\sigma_{\mathrm{AH}} \propto \sigma_{xx}^{1.6}$, that is coincident with a recent theory. This relation is valid over five decades of $\sigma_{xx}$ irrespective of metallic or hopping conduction.

---

*E-mail address: fukumura@imr.tohoku.ac.jp

Anomalous Hall effect (AHE) is generally observed in ferromagnetic metals and semiconductors, and empirically expressed as $\rho_H = R_0 H + R_S M$ ($\rho_H$: Hall resistivity; $R_0$: normal Hall coefficient; $H$: magnetic field; $R_S$: anomalous Hall coefficient; $M$: magnetization).[1)] Microscopic theory of AHE has been of long debate for half a century, and is being developed quite recently to explain AHE in various ferromagnetic metals quantitatively, in which the Berry phase of Bloch wave function plays a crucial role.[2)] For general consideration of AHE, more comprehensive treatment is needed than the well known skew scattering[3)] and side jump mechanisms.[4)] The former and the latter are respectively known to yield in the relations of $\rho_H \propto \rho_{xx}$ and $\rho_H \propto \rho_{xx}^2$, where $\rho$ stands for resistivity. In a recent theory for multiband ferromagnetic metals with dilute impurities, the power law dependence of the anomalous Hall conductivity $\sigma_{AH}$ on the conductivity $\sigma_{xx}$ was shown to have an extrinsic-to-intrinsic crossover with different exponents at a certain value of conductivity.[5)] The extrinsic skew scattering mechanism ($\sigma_{AH} \propto \sigma_{xx}^1$) appears in the clean limit with higher conductivity, whereas the intrinsic contribution becomes dominant with lowering the conductivity. In the dirty limit, the intrinsic contribution is subject to the damping due to impurities yielding in the relation of $\sigma_{AH} \propto \sigma_{xx}^{1.6}$, where the exponent is close to that in normal Hall effect of quantum Hall insulator.[6)] This theory is based on the use of Bloch wave function assuming the metallic conduction, hence the result is valid only for ferromagnetic metals in principle. Indeed, recent experiments for ferromagnetic metals supported the theory in the regime of high conductivity.[7)]

$(Ti,Co)O_2$, both in rutile and anatase phases, is a high temperature ferromagnetic semiconductor.[8-10)] Both phases of $(Ti,Co)O_2$ show AHE, where the rutile $(Ti,Co)O_2$ has one decade larger anomalous part of $\rho_H$ than the anatase $(Ti,Co)O_2$.[11,12)] Previously, we reported that the rutile $(Ti,Co)O_2$ show a scaling relation of AHE: the anomalous Hall conductivity approximately scales with the conductivity as $\sigma_{AH} \propto \sigma_{xx}^{1.5\text{-}1.7}$ irrespective of the Co content, the measurement temperature, or the value of conductivity.[11)] In this rutile $(Ti,Co)O_2$, the conductivity could be varied for several decades by the contents of oxygen deficiencies as donors [$10^{-2}\ \Omega^{-1}\cdot cm^{-1} \leq \sigma_{xx}$ (300 K) $\leq 10^{2}\ \Omega^{-1}\cdot cm^{-1}$], maintaining the hopping conduction as a result of freezing out of the charge carriers at low temperature. Recently, we observed that the anatase $(Ti,Co)O_2$, with the much higher mobility than the rutile $(Ti,Co)O_2$, merges into the same scaling relation in spite of the metallic conduction [$\sigma_{xx}$ (300 K) $\geq 10^{2}\ \Omega^{-1}\cdot cm^{-1}$],[12] implying the universal feature of the scaling relation. In this study, we perform a comprehensive survey of the existing data for ferromagnetic oxide semiconductors $(Ti,Co)O_2$, Mn-doped III-V semiconductors such as (Ga,Mn)As, correlated electron ferromagnetic oxides such as $(La,Sr)MnO_3$, transition metal silicides such as (Fe,Co)Si, and metals such as Fe and Co. Some of them show metallic conduction and the others show hopping conduction. Surprisingly, the scaling relation is confirmed to be present for many of these compounds irrespective of the metallic or hopping conduction.

Various ferromagnetic semiconductors and metals are listed in Table I available in literatures.[7,11-29)] The measurement temperatures of $\sigma_{AH}$ and $\sigma_{xx}$ and the Curie temperatures ($T_C$) are also shown. The measurement temperatures were sometimes higher than $T_C$ since the nonlinear dependence of Hall resistivity in the magnetic field was observed above $T_C$.[30)] Whether metallic or hopping conduction for each compound is determined by the dependence of resistivity on temperature.

Figure 1 shows $|\sigma_{AH}|$ vs $\sigma_{xx}$ relationship for each compound listed in Table I. For $(Ti,Co)O_2$, the anomalous Hall conductivity was approximately deduced to be $\rho_H/\rho_{xx}^2$ since $\rho_H \ll \rho_{xx}$, where the contribution of the normal Hall part was excluded.[11,12)] For the other compounds, the conductivity and the anomalous Hall conductivity were respectively deduced as $\rho_{xx}/(\rho_{xx}^2 + \rho_H^2)$ and $\rho_H/(\rho_{xx}^2 + \rho_H^2)$ if underived in the original literatures, where $\rho_H$ was evaluated as linearly extrapolated value to zero magnetic field from the saturated value in high magnetic field. For discussion of ferromagnetic metals with high conductivity such as Fe, see ref. 8. Here, let us discuss on the ferromagnetic metals and semiconductors with relatively low conductivity $\sigma_{xx} \leq 10^4\ \Omega^{-1}\cdot\mathrm{cm}^{-1}$.

The relationship for the rutile and anatase $(Ti,Co)O_2$ after different research groups follows a scaling line $\sigma_{AH} \propto \sigma_{xx}^{1.6}$ for five decades of the conductivity.[11-14,31] The scaling relation agrees with the theory for the regime of low conductivity (ref. 5), although the conductivity of $(Ti,Co)O_2$ extends to much lower values.

For Mn-doped III-V semiconductors such as (Ga,Mn)As, the relationship shows the similar behavior. (Ga,Mn)As, (In,Mn)As, (Ga,Mn)Sb, and (In,Mn)Sb follow another scaling line $\sigma_{AH} \propto \sigma_{xx}^{1.6}$ over five decades of the conductivity,[15-22] irrespective of the Mn contents and the measurement temperatures. In these compounds, Mn serves as dopant of both holes and spins, hence the conduction state can be varied from metallic to hopping conduction. As can be seen in Fig. 1, the scaling relation is independent of the metallic and hopping conduction. It is noted that $|\sigma_{AH}|$ for Mn-doped III-V semiconductors is about three decades larger than that for $(Ti,Co)O_2$.

For the higher regime of the conductivity, various correlated electron ferromagnetic oxides locate. $(La,A)MnO_3$ is usually ferromagnetic metal with double exchange mechanism, where alkaline earth element *A* serves as hole dopants, and the contents of *A* also affect the magnetic phases. For $(La,A)MnO_3$, the scaling relation is nearly extrapolated from $(Ti,Co)O_2$.[23,24] In $(La,Sr)CoO_3$, the conduction state depends on the Sr contents: varying from hopping to metallic conduction with increasing Sr contents.[25] The scaling relation is also unchanged for the metallic or hopping conduction, as observed for $(Ti,Co)O_2$ and Mn-doped III-V semiconductors. Higher conducting oxides with $\sigma_{xx} \sim 10^{3\text{-}4}\ \Omega^{-1}\cdot\mathrm{cm}^{-1}$ such as $SrRuO_3$, $(Sr,Ca)RuO_3$, $Sr_2FeMoO_6$, and $Nd_2Mo_2O_7$ approximately show the similar behavior of the scaling relation.[7,26-28] Transition metal silicides (Fe,Co)Si and (Fe,Mn)Si also locate around the above transition metal oxides.[29)]

Theoretically, the magnitude of anomalous Hall conductivity is explained to be determined by the degree of resonance caused by the location of Fermi level around an anticrossing of band dispersion.[5)] Considering that the band structure might take the principal role, it is noted that several compounds show the similar scaling relation with the similar magnitude of the anomalous Hall conductivity irrespective of the different crystal structures, the dense or dilute magnetic ion systems, and the metallic or hopping conduction. This theory assumes the metallic conduction, however, the scaling relation is evidenced for the regime of hopping conduction as demonstrated experimentally in Fig.1.

In summary, a scaling relation of the anomalous Hall effect recently found for $(Ti,Co)O_2$, $\sigma_{AH} \propto \sigma_{xx}^{1.6}$, seems to be universal over five decades of the conductivity.($\sigma_{xx} \leq 10^4\ \Omega^{-1}\cdot cm^{-1}$) in various ferromagnetic semiconductors and metals, irrespective of the metallic or hopping conduction implying the universal feature of the scaling relation.

**Acknowledgements**

The authors acknowledge for N. Nagaosa, S. Onoda, H. Ohno, and F. Matsukura for fruitful discussions. This work was supported by the Ministry of Education, Culture, Sports, Science and Technology (MEXT), Japan for Young Scientists (A19686021) and for Scientific Research on Priority Areas (16076205), New Energy and Industrial Technology Development Organization, Industrial Technology Research Grant Program (05A24020d), and Tokyo Ohka Foundation for the Promotion of Science and Technology.

**References**

1) C. L. Chien and C. R. Westgate: *The Hall Effect and Its Applications* (Plenum, New York, 1979).
2) N. Nagaosa: J. Phys. Soc. Jpn. **75** (2006) 042001.
3) J. Smit: Physica (Amsterdam) **24** (1958) 39.
4) L. Berger: Phys. Rev. B **2** (1970) 4559.
5) S. Onoda, N. Sugimoto, and N. Nagaosa: Phys. Rev. Lett. **97** (2006) 126602.
6) L. P. Pryadko and A. Auerbach: Phys. Rev. Lett. **82** (1999) 1253.
7) T. Miyasato, N. Abe, T. Fujii, A. Asamitsu, S. Onoda, Y. Onose, N. Nagaosa, and Y. Tokura: cond-mat/0610324.
8) Y. Matsumoto, M. Murakami, T. Shono, T. Hasegawa, T. Fukumura, M. Kawasaki, P. Ahmet, T. Chikyow, S. Koshihara, and H. Koinuma: Science **291** (2001) 854.
9) Y. Matsumoto, R. Takahashi, M. Murakami, T. Koida, X.-J. Fan, T. Hasegawa, T. Fukumura, M. Kawasaki, S. Koshihara, and H. Koinuma: Jpn. J. Appl. Phys. **40** (2001) L1204.
10) T. Fukumura, H. Toyosaki, and Y. Yamada: Semicond. Sci. Technol. **20** (2005) S103.
11) H. Toyosaki, T. Fukumura, Y. Yamada, K. Nakajima, T. Chikyow, T. Hasegawa, H. Koinuma, and M. Kawasaki: Nat. Mater. **3** (2004) 221.
12) K. Ueno, T. Fukumura, H. Toyosaki, M. Nakano, and M. Kawasaki: Appl. Phys. Lett. **90** (2007) 072103.
13) J. S. Higgins, S. R. Shinde, S. B. Ogale, T. Venkatesan, and R. L. Greene: Phys. Rev. B **69** (2004) 073201.
14) T. Hitosugi, G. Kinoda, Y. Yamamoto, Y. Furubayashi, K. Inaba, Y. Hirose, K. Nakajima, T. Chikyow, T. Shimada, and T. Hasegawa: J. Appl. Phys. **99** (2006) 08M121.
15) F. Matsukura, H. Ohno, A. Shen, and Y. Sugawara: Phys. Rev. B **57** (1998) R2037.
16) K. W. Edmonds, R. P. Campion, K.-Y. Wang, A. C. Neumann, B. L. Gallagher, C. T. Foxon, and P. C. Main: J. Appl. Phys. **93** (2003) 6787.
17) S. U. Yuldashev, H. C. Jeon, H. S. Im, T. W. Kang, S. H. Lee, and J. K. Furdyna: Phys. Rev. B **70** (2004) 193203.
18) D. Chiba, Y. Nishitani, F. Matsukura, and H. Ohno: Appl. Phys. Lett. **90** (2007) 122503.
19) H. Ohno, H. Munekata, T. Penny, S. von Molnar, and L. L. Chang: Phys. Rev. Lett. **68**

(1992) 2664.
20) A. Oiwa, A. Endo, S. Katsumoto, Y. Iye, and H. Ohno: Phys. Rev. B **59** (1999) 5826.
21) F. Matsukura, E. Abe, and H. Ohno: J. Appl. Phys. **87** (2000) 6442.
22) T. Wojtowicz, W. L. Lim, X. Liu, G. Cywinski, M. Kutrowski, L.V. Titova, K. Yee, M. Dobrowolska, J. K. Furdyna, K. M. Yu, W. Walukiewicz, G. B. Kim, M. Cheon, X. Chen, S. M. Wang, H. Luo, I. Vurgaftman, and J. R. Meyer: Physica E **20** (2004) 325.
23) Y. Lyanda-Geller, S. H. Chun, M. B. Salamon, P. M. Goldbart, P. D. Han, Y. Tomioka, A. Asamitsu, and Y. Tokura: Phys. Rev. B **63** (2001) 184426.
24) P. Matl, N. P. Ong, Y. F. Yan, Y. W. Li, D. Studebaker, T. Baum, and G. Doubinina: Phys. Rev. B **57** (1998) 10248.
25) Y. Onose and Y. Tokura: Phys. Rev. B **73** (2006) 174421.
26) R. Mathieu, A. Asamitsu, H. Yamada, K. S. Takahashi, M. Kawasaki, Z. Fang, N. Nagaosa, and Y. Tokura: Phys. Rev. Lett. **93** (2004) 016602.
27) Y. Tomioka, T. Okuda, Y. Okimoto, R. Kumai, K.-I. Kobayashi, and Y. Tokura: Phys. Rev. B **61** (2000) 422.
28) Y. Taguchi, Y. Oohara, H. Yoshizawa, N. Nagaosa, and Y. Tokura: Science **291** (2001) 2573.
29) N. Manyala, Y. Sidis, J. F. Ditusa, G. Aeppli, D. P. Young, and Z. Fisk: Nat. Mater. **3** (2004) 255.
30) Several data near $T_C$ accompany the significant change in the magnetization. However, influence of such change might be hindered in Fig. 1 due to its logarithmic scale.
31) Higgins *et al.* attributed the presence of the anomalous Hall effect to the precipitation of ferromagnetic Co metal in the $TiO_2$ matrix,[13)] however, the very low Co content (2%) could not yield in the observable AHE. Instead, their result can be interpreted as the appearance of AHE in $(Ti,Co)O_2$.

**Figure captions**

Fig. 1. Relationship between the magnitude of anomalous Hall conductivity ($\sigma_{AH}$) and the conductivity ($\sigma_{xx}$) for various ferromagnetic semiconductors and metals in Table I. Open and solid symbols denote metallic and hopping conductions defined in Table I, respectively. Dotted lines represent $\sigma_{AH} \propto \sigma_{xx}^{1.6}$ lines.

Table I. List of ferromagnetic semiconductors and metals. The measured temperature ($T$), the Curie temperature ($T_C$), and the conduction states are shown. a- and r- $TiO_2$ denote anatase and rutile $TiO_2$. Missing values are not described in original literatures.

| Compounds | $T$ (K) | $T_C$ (K) | Conduction states | References |
|---|---|---|---|---|
| r-$Ti_{1-x}Co_xO_2$ | 100-300 | > 400 | hopping | 11 |
| r-$Ti_{0.98}Co_{0.02}O_2$ | 200-300 | > 400 | hopping | 13 |
| a-$Ti_{0.95}Co_{0.05}O_2$ | 10-350 | > 400 | metallic | 12 |
| a-$Ti_{0.92}Co_{0.05}Nb_{0.03}O_2$ | 200 | > 400 | metallic | 14 |
| $Ga_{0.947}Mn_{0.053}As$ | 2-150 | 110 | metallic | 15 |
| $Ga_{0.94}Mn_{0.06}As$ | 0.4 | 132 | metallic | 16 |
| $Ga_{0.915}Mn_{0.085}As$ | 10-60 | 0-30 | hopping | 17 |
| $Ga_{1-x}Mn_xAs$ | 10-160 | 120-145 | metallic | 18 |
| $In_{0.987}Mn_{0.013}As$ | 3.5 | 7.5 | hopping | 19 |
| $In_{1-x}Mn_xAs$ | 4.2 | 35 | metallic | 20 |
| $Ga_{0.977}Mn_{0.023}Sb$ | 1.5-30 | 25 | metallic | 21 |
| $In_{0.98}Mn_{0.02}Sb$ | 1.5-12 | 7 | metallic | 22 |
| $La_{0.7}Ca_{0.3}MnO_3$ | 10-300 | 216 | metallic | 23 |
| $La_{0.7}Ca_{0.3}MnO_3$ | 250-320 | 265 | metallic | 24 |
| $La_{0.67}(Ca,Pb)_{0.33}MnO_3$ | 100-350 | 285 | metallic | 23 |
| $La_{0.7}Sr_{0.3}MnO_3$ | 10-400 | 362 | metallic | 23 |
| $La_{1-x}Sr_xCoO_3$ | 40-300 | 120-230 | hopping, metallic | 25 |
| $SrRuO_3$ single crystal | - | 160 | metallic | 7 |
| $Sr_{1-x}Ca_xRuO_3$ | 2-160 | 70-150 | metallic | 26 |
| $Sr_2FeMoO_6$ | 5 | 420 | metallic | 27 |
| $Nd_2Mo_2O_7$ | 2-100 | 89 | metallic | 28 |
| $Fe_{1-x}Mn_xSi$ | 5 | 0-30 | metallic | 29 |
| $Fe_{1-x}Co_xSi$ | 5 | 10-47 | metallic | 29 |
| Fe single crystal | - | - | metallic | 7 |
| Fe film | - | - | metallic | 7 |
| Co film | - | - | metallic | 7 |
| Ni film | - | - | metallic | 7 |
| Gd film | - | - | metallic | 7 |

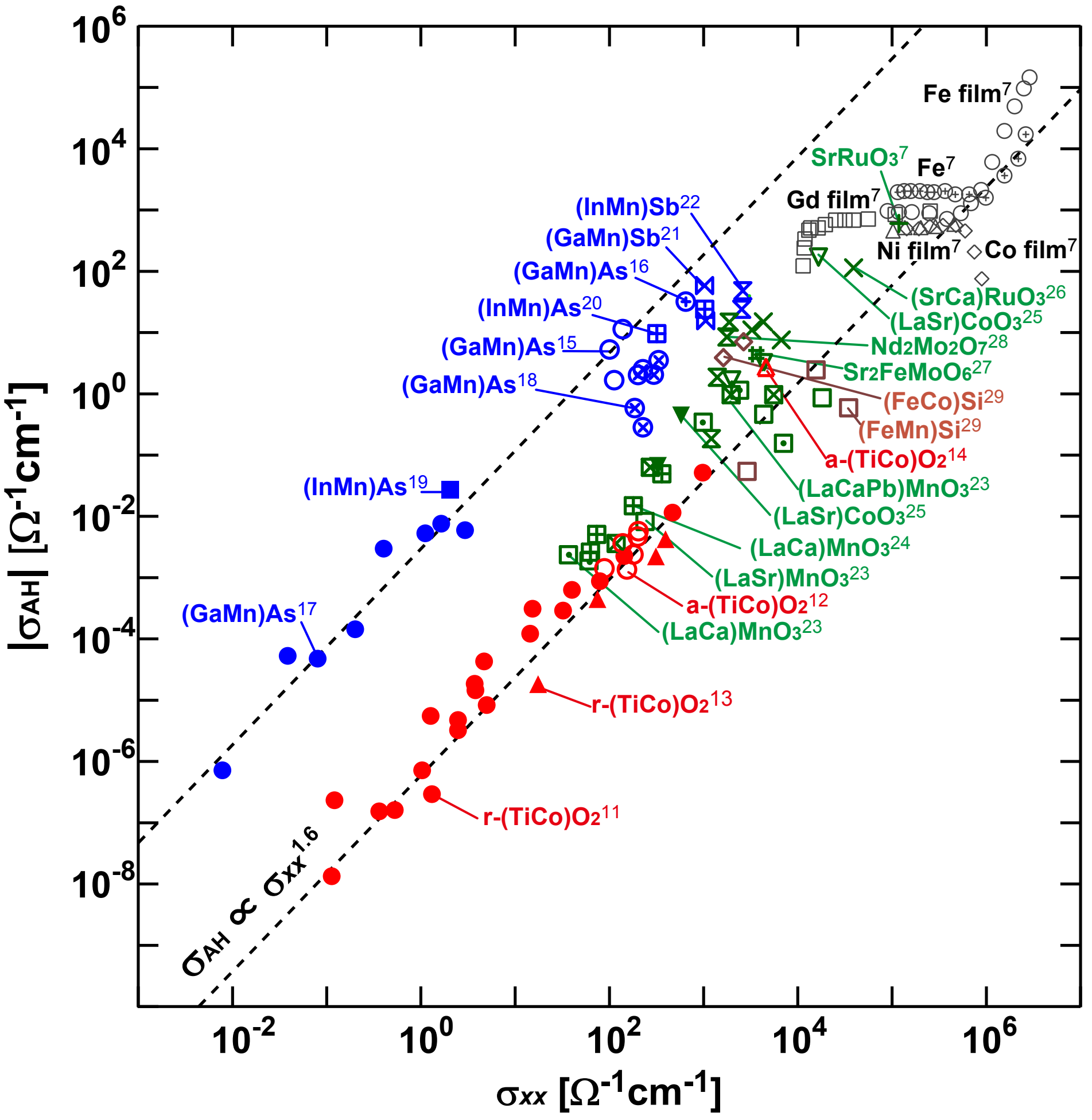

Figure 1 T. Fukumura et al.